\documentclass[aps,prl,reprint,superscriptaddress,amsmath,notitle]{revtex4-2}
\usepackage{graphicx}
\usepackage{enumerate}
\usepackage{refcount}
\usepackage{amssymb}
\usepackage{dsfont}
\usepackage{hyperref}
\usepackage[usenames,dvipsnames]{color}
\usepackage[normalem]{ulem}
\usepackage{physics}
\usepackage[T1]{fontenc}
\usepackage[utf8x]{inputenc}
\usepackage{pdfpages}

\makeatletter
\AtBeginDocument{\let\LS@rot\@undefined}
\makeatother

\hypersetup{pdfstartview=FitH,pdfpagemode=UseOutlines,colorlinks,
citecolor=blue,linkcolor=blue,urlcolor=blue}

\newcommand{\phd}{\vphantom{\dagger}}

\begin{document}

\title{Detecting Fractional Chern Insulators in Optical Lattices through Quantized Displacement}

\author{Johannes Motruk}
    \affiliation{Department of Physics, University of California, Berkeley, California 94720, USA}
	\affiliation{Materials Sciences Division, Lawrence Berkeley National Laboratory, Berkeley, California, 94720, USA}
\author{Ilyoun Na}
	\affiliation{Department of Physics, University of California, Berkeley, California 94720, USA}

\date{\today}

\begin{abstract}
The realization of interacting topological states of matter such as fractional Chern insulators (FCIs) in cold atom systems has recently come within experimental reach due to the engineering of optical lattices  with synthetic gauge fields providing the required topological band structures. However, detecting their occurrence might prove difficult since transport measurements akin to those in solid state systems are challenging to perform in cold atom setups and alternatives have to be found. We show that for a $\nu= 1/2$ FCI state realized in the lowest band of a Harper-Hofstadter model of interacting bosons confined by a harmonic trapping potential, the fractionally quantized Hall conductivity $\sigma_{xy}$ can be accurately determined by the displacement of the atomic cloud under the action of a constant force which provides a suitable experimentally measurable signal for detecting the topological nature of the state. Using matrix-product state algorithms, we show that, in both cylinder and square geometries, the movement of the particle cloud in time under the application of a constant force field on top of the confining potential is proportional to $\sigma_{xy}$ for an extended range of field strengths.

\end{abstract}

\maketitle


\textit{Introduction}.---Ultracold atoms in optical lattices have recently become a fruitful field for the realization of topological states of matter~\cite{Jotzu2014,Aidelsburger2015,Wu2016,Goldman2016,Song2018,Zhang2018,Cooper2019,Song2019,Wintersperger2020}. 
The engineering of artificial gauge fields~\cite{Dalibard2011,Goldman2014b} 
has enabled the creation of band structures with nontrivial topology~\cite{Aidelsburger2013,Miyake2013,Jotzu2014} and their nonzero Chern numbers~\cite{Aidelsburger2015,Sun2018,Tarnowski2019,Asteria2019,Genkina2019,Chalopin2020} and chiral edge states~\cite{Mancini2015,Stuhl2015} have been detected in the laboratory. While those experiments focused on noninteracting systems, first steps have been taken towards the observation of strongly correlated topological states~\cite{Tai2017} which can soon be expected to lead to the experimental realization of fractional Chern insulators (FCIs)~\cite{Kol1993,Bergholtz2013,Parameswaran2013}, lattice versions of the fractional quantum Hall effect~\cite{Tsui1982}. The hallmark signature of these states is their quantized Hall conductivity which, in the solid state, is conventionally measured by transport experiments that, however, cannot be straightforwardly adopted into the cold atom setting. To this end, theoretical studies have proposed to probe their topological properties through their characteristic edge states~\cite{Kjaell2012,Goldman2013,He2017,Dong2018}, fractionalization of quasiparticles~\cite{Grusdt2016, Raciunas2018, Umucalilar2018,Macaluso2020}, or circular dichroism~\cite{Repellin2019}.
Yet, it is still unclear which protocol is experimentally most feasible and it remains a crucial challenge to determine unambiguous signatures for the topological nature of a putative FCI state in an optical lattice.

\begin{figure}[t]
    \centering
    \includegraphics[width=\columnwidth]{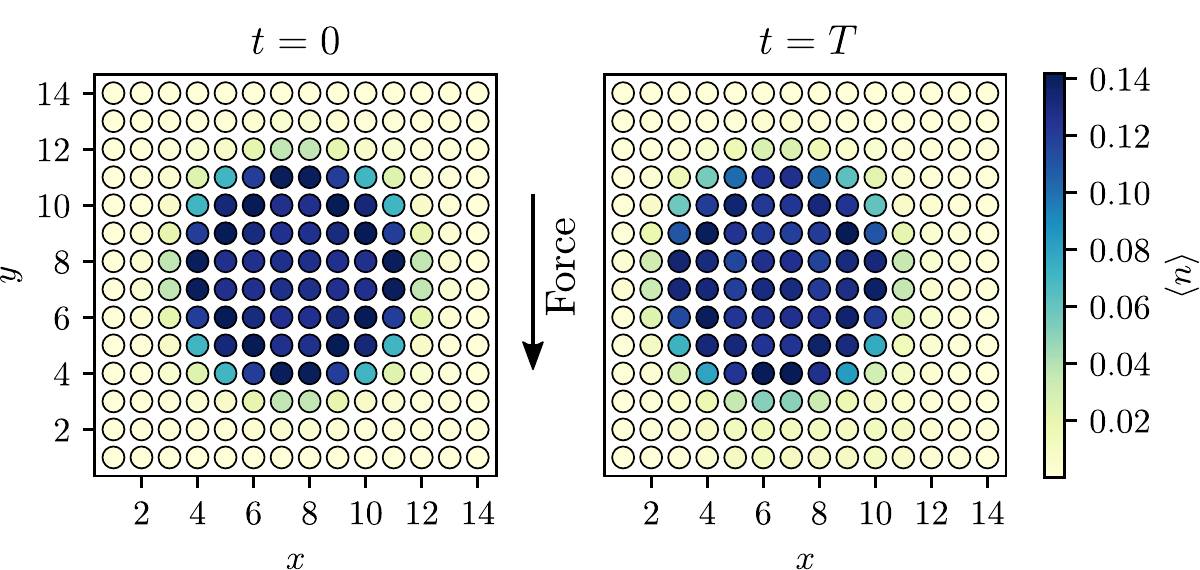}
    \caption{Quantized displacement of an FCI cloud in a harmonic trap shown by the particle density $\expval{n}$ at time $t=0$ and $t=T=9$. A force in the $-y$ direction leads to a drift of the particles in the $-x$ direction proportional to the Hall conductivity $\sigma_{xy}=1/2$. The force field is of strength $E_y=-\pi/18 \approx 0.175$ and the confinement $k_x=k_y=0.024$, see Eq.~\eqref{eq:ham}. }
    \label{fig:displace}
\end{figure}

In this Letter, we propose that the topology of an FCI state can be detected by a drift of the atomic cloud proportional to its Hall conductivity $\sigma_{xy}$ under a constant force field akin to the noninteracting case~\cite{Goldman2013a,Aidelsburger2015,Price2016}, see Fig.~\ref{fig:displace}. We demonstrate this behavior by performing simulations of the response of an FCI particle cloud confined by a harmonic potential on cylinder and open systems and show that $\sigma_{xy}$ can be accurately determined even in systems containing a number of $\sim 10$ particles.

\textit{Model and method}.---We consider the interacting Harper-Hofstadter model~\cite{Hofstadter1976} for bosons on a square lattice on a cylinder and square geometry with Hamiltonian $H = H_0 + H_U$ and
\begin{align} 
H_0 =& -J \sum_{x,y} \left( b^{\dag}_{x+1,y}b^{\phd}_{x,y} + e^{i \phi} b^{\dag}_{x,y+1}b^{\phd}_{x,y} + {\rm H.c.} \right)  \label{eq:ham} \\ 
& +\sum_{x,y} \left[ \frac{k_x}{2} (x-x_0)^{2} + \frac{k_y}{2} (y-y_0)^{2} + E_y y \right] n_{x,y}, \nonumber
\end{align}
where $b^{\dag}_{x,y}$ ($b^{\phd}_{x,y}$) creates (annihilates) a boson and $n_{x,y} = b^{\dag}_{x,y} b^{\phd}_{x,y} $ measures the occupation number at site $(x,y)$. The phase $\phi = \pi x/2 - \phi_{\rm ext}/L_y$ generates both a flux of $\pi/2$ per square plaquette and $\phi_{\rm ext}$ through the cylinder. The centers of the confining potentials are given by $x_0(y_0) = (L_{x(y)}-1)/2$ and $E_y$ denotes the electric field in the $y$ direction in the square geometry. The interacting term reads $H_U = (U/2)\sum_{x,y} (n_{x,y}-1)n_{x,y}$, but we consider the hardcore boson limit $U/J \to \infty$ and set $J=1$ so that all energy values are given in units of $1/J$. This model exhibits four single-particle bands and has been shown to host a $\nu=1/2$ Laughlin state~\cite{Laughlin1983,Kalmeyer1987} at half filling of the lowest Chern number $C=1$ band for both large and infinite $U$ in several studies~\cite{Sorensen2005,Hafezi2007,Kjaell2012,He2017,Motruk2017,Rosson2019}.
To investigate the interacting model, we compute the ground state using the density matrix renormalization group algorithm~\cite{White1992} and simulate the time evolution of the system including an electric field to extract $\sigma_{xy}$ with the algorithm introduced in Ref.~\cite{Zaletel2015}. Note that we use the term electric field in analogy to the $U(1)$ gauge theory of electromagnetism, yet we do not refer to a physical electric field and the bosonic particles do not need to have---and in experiments with cold atoms---will not have an electric charge.

\begin{figure}[t]
    \centering
    \includegraphics[width=\columnwidth]{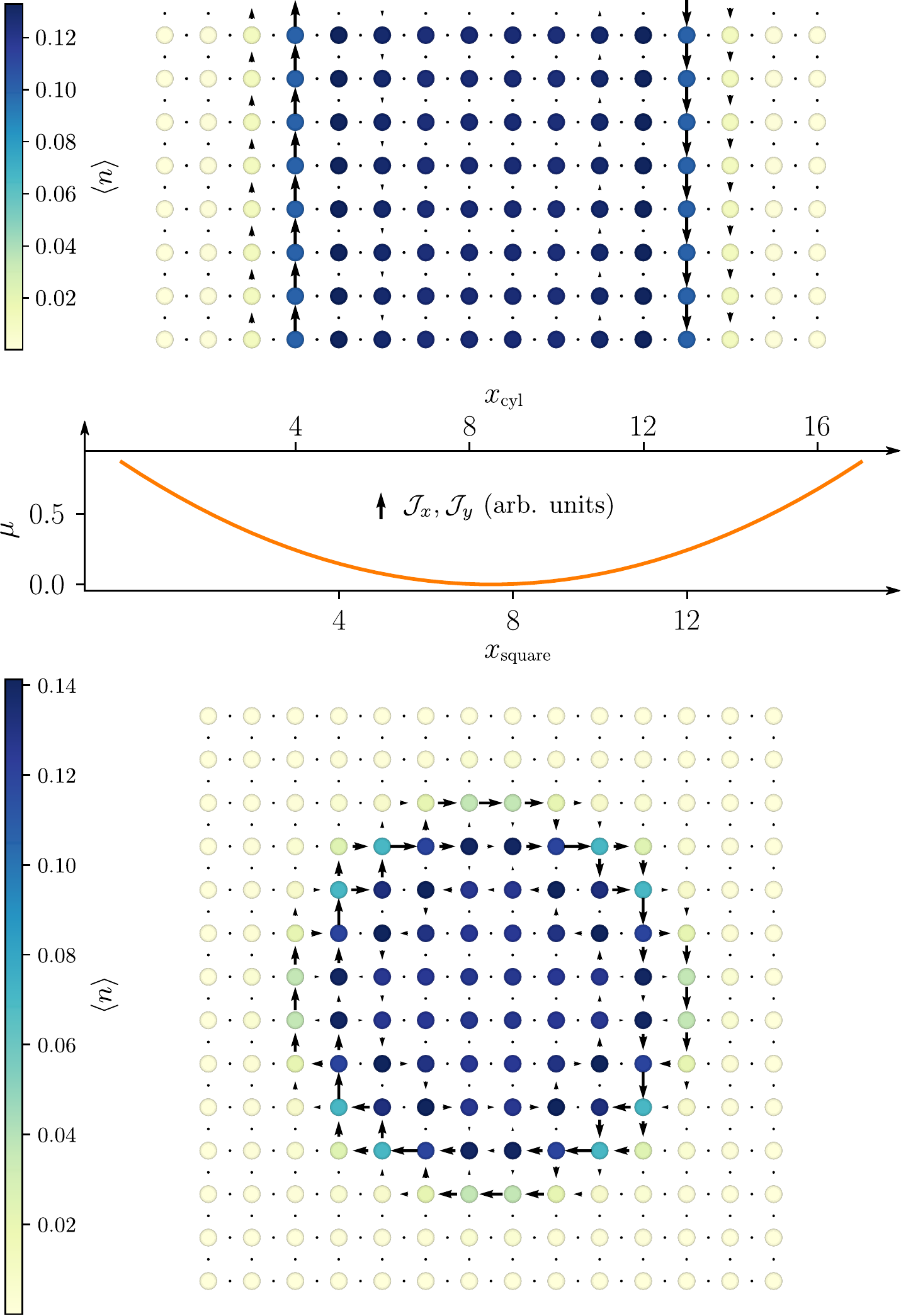}
    \caption{Particle and current density of FCI states for cylinder (upper part, $L_y=8$, $k_x=0.024$, $N=10$) and square geometry (lower part, $L_x \times L_y = 14 \times 14$, $k_x=k_y=0.024$, $N=8$). Both systems exhibit extended central regions of $\expval{n} \approx 1/8$ with chiral edge currents propagating around.
    The chemical potential $\mu$ of the harmonic trap is plotted in between for orientation and applies to both cylinder and square cases in the $x$ direction with the respective coordinate axis.}
    \label{fig:dens_curr}
\end{figure}

\textit{Quantized displacement}.---We focus on four geometries: three cylinders of length $L_x=16$ and width $L_y=4,6,8$ with $5,7$ and $10$ particles and an open system with $L_x \times L_y = 14 \times 14$ and $8$ particles. As confinement strength, we choose $k_x=0.024$ on the cylinders and $k_x=k_y=0.024$ for the open system which leads to extended regions of filling $\expval{n} \approx 0.125$ in the centers of the system compatible with half filling of the lowest band (see Fig.~\ref{fig:dens_curr}). We present the effect of the trap strength on the density distribution in the cylinder case in more detail in \cite{SuppMat}.

\begin{figure}[t]
    \centering
    \includegraphics[width=\columnwidth]{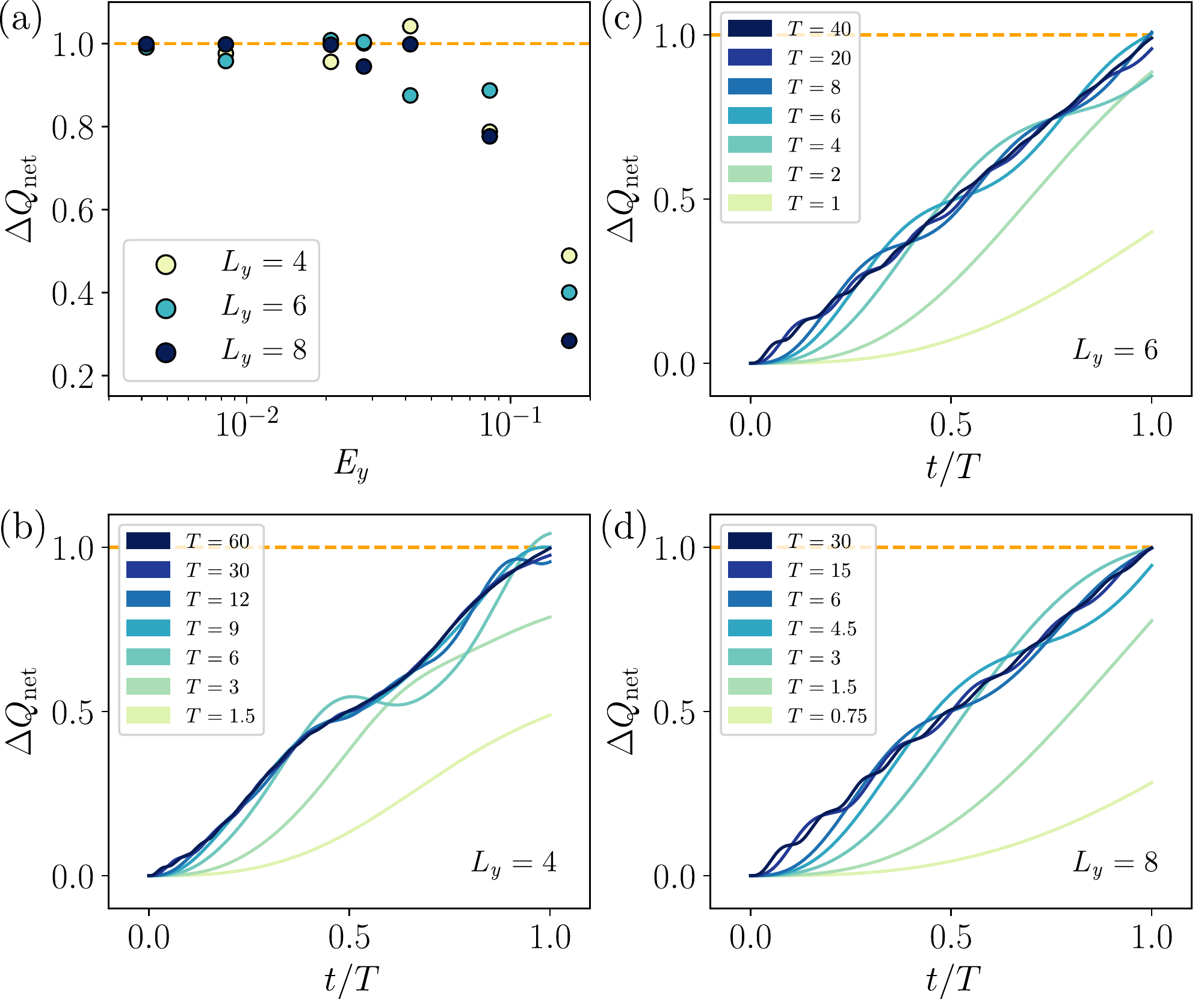}
    \caption{Behavior of the averaged pumped charge on the cylinder system $\Delta Q_{\rm net}(t)$ (a) at $t=T$ after a flux insertion of $4\pi$ as a function of the field strength $E_y$ for $L_{y}=4,6,8$. The pumped charge converges to unity for small field strengths $E_y$ (large flux insertion time $T$) indicating a Hall conductivity of $\sigma_{xy}=1/2$. (b)-(d) Full time evolution of $\Delta Q_{\rm net}$ for different ramp times $T$ at each circumference (b) $L_{y}=4$, (c) $L_{y}=6$, and (d) $L_{y}=8$. The times for different $L_y$ have been chosen such that they lead to the same field strengths for all cylinder circumferences. }
    \label{fig:pump}
\end{figure}

We study how the ground states evolve in time when the particles are subject to an additional constant force and first focus on the cylinder system. Therefore, we dynamically tune the external flux $\phi_{\rm ext}$ through the cylinder from time $t=0$ to $t=T$ as $\phi_{\rm ext} (t) = 4\pi t/T$. This varying flux induces an electric field around the circumference of the cylinder of the form $\vec E =-d{\vec A}/{dt}=-(4\pi / (T L_{y})) \hat y$.
If the central region forms a $\nu=1/2$ Laughlin state, we expect a bulk Hall conductivity of $\sigma_{xy}=1/2$ leading to a current response of $j_x = \sigma_{xy} E_y$.  Initially, the cloud spreads over approximately $8 \times L_{y}$ sites with tails of lower particle density on the edges. In order to test the bulk response, we monitor the quantity $\Delta Q_{\rm net}$ defined below. The charge that has flown at time $t$ through a cut between the sites at $x=i$ and $x=i-1$ is given by
$\Delta Q_{i \to i-1} (t) =\sum_{x=1}^{i-1} \sum_{y=1}^{L_{y}} ( \expval{n_{x,y}(t)} - \expval{n_{x,y}(0)} )$.
Then, we take the average of this quantity over the particle cloud at the center of the system and define
\begin{equation}
\Delta Q_{\rm net} (t) = \frac{1}{8} \sum_{i=5}^{12} \Delta Q_{i \to i-1} (t). \label{eq:pump}
\end{equation} 
With $T\to \infty$, this quantity should approach unity when two flux quanta ($4\pi$) have been inserted through the cylinder according to Laughlin's gauge argument~\cite{Laughlin1981}. 
Note that we have found this quantity to be more accurate than the center of mass of the particles for determining the Hall conductivity as the effects of the tails of the particle density around the edges without quantized $\sigma_{xy}$ can be excluded in this way.
We show the results for the final value $\Delta Q_{\rm net}(T)$ for the different circumferences in Fig.~\ref{fig:pump}(a) where the ramping times $T$ are chosen such that the flux insertion induces electric fields of equal strengths for the respective $L_y$ as $E_y \propto 1/(TL_y)$. The value of $\Delta Q_{\rm net}(T)$ clearly approaches unity as the electric field is decreased ($T$ increased). Especially with increasing $L_y$, the quantity seems to converge earlier and in a more controlled manner which can be understood by looking at the behavior of $\Delta Q_{\rm net}(t)$ over the full evolution time shown in Figs.~\ref{fig:pump}(b)--\ref{fig:pump}(d). For small times, the charge pumping is clearly nonadiabatic resulting in a final value below unity. With increasing $T$, however, the $\Delta Q_{\rm net}$-$t/T$ curve approaches a line with slope one exhibiting oscillations around it. 
These oscillations become weaker with increasing evolution time as $\propto 1/T \propto E_y  L_y$ resulting in a quicker convergence of $\Delta Q_{\rm net}(T)$ as a function of $E_y$ to unity at larger circumferences.

\begin{figure}[t]
    \centering
    \includegraphics[width=\columnwidth]{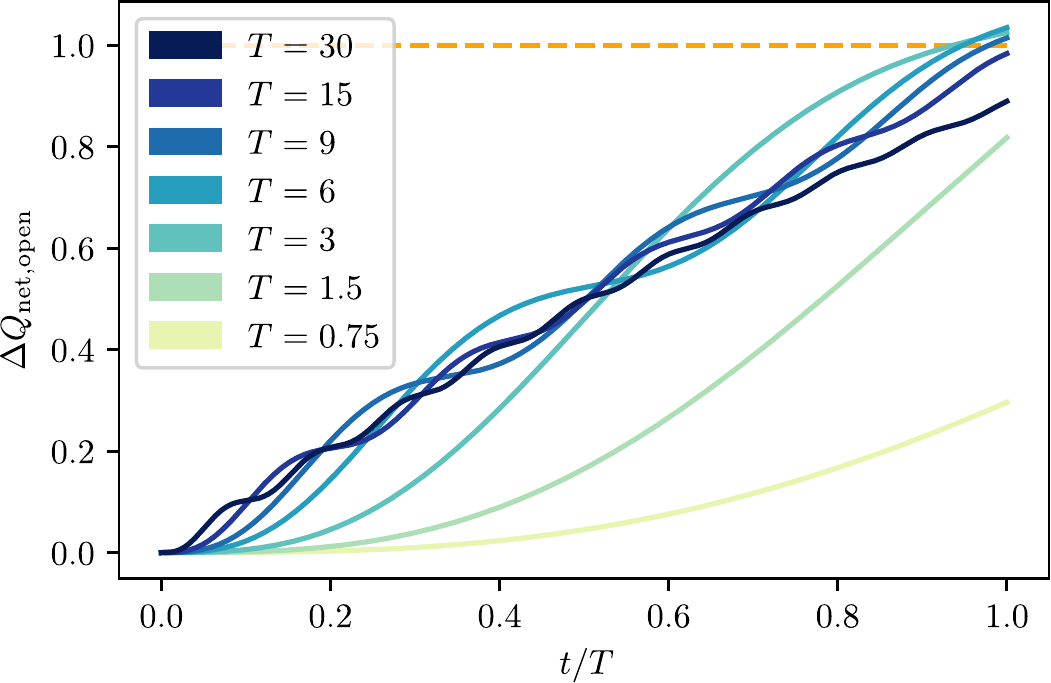}
    \caption{Time evolution of the pumped charge $\Delta Q_{\rm net, open} $ in the system with square geometry and harmonic confinement in both the $x$ and $y$ direction. The behavior is qualitatively similar to the $L_y=8$ case from Fig.~\ref{fig:pump}(d), but the total pumped charge decreases for larger $T$ due to the competition between confining and field-creating potential (see main text).}
    \label{fig:pump_open}
\end{figure}

Now, we turn to the time evolution in the square system with open boundary conditions. The electric field in this case is simply created by a static potential $\Phi_y = E_y y$ and we set its strength equal to the one for the $L_y=8$ cylinder, i.e. $E_y = -\pi/(2T)$. Since the particle cloud only extends over eight sites in the $y$ direction on columns $x~=~\{6, \ldots, 9\}$ (see Fig.~\ref{fig:displace}), we average $\Delta Q$ only over these and compute it as
$\Delta Q_{\rm net, open} (t) = \frac{1}{4} \sum_{i=6}^{9} \Delta Q_{i \to i-1} (t)$
leading to the behavior shown in Fig.~\ref{fig:pump_open}. For small evolution times $T$ (large field strengths), it is qualitatively very similar to the $L_y=8$ cylinder case, however, the value at $t=T$ decreases for larger times. We attribute this to the fact that at small values of the field, the confining potential gradient in the area of the cloud, especially at the edges, becomes comparable to the potential gradient creating the electric field. 

\textit{Comparison to Chern insulator of free fermions}.---While the quantization of the total charge pumping $\Delta Q_{\rm net} (T) $ on the cylinder becomes ever more accurate with increasing $T$, the value in the experimentally more realistic square geometry decreases for smaller field strengths.
Larger system sizes allow for weaker trap strengths $k_x$ and $k_y$, however, are not feasible to simulate for the interacting system. Therefore, we investigate whether the displacement behavior is similar in an integer Chern insulator (CI) of free fermions and how it is influenced by the system size. 

To this end, we consider the Hamiltonian from Eq.~\eqref{eq:ham} for noninteracting fermions instead of hardcore bosons  and compute the displacement on a cylinder and open system of comparable size to the interacting case. An integer CI in this system forms when the lowest band is completely filled with fermions, corresponding to a particle density of $\expval{n}=1/4$.
\begin{figure}[t]
    \centering
    \includegraphics[width=\columnwidth]{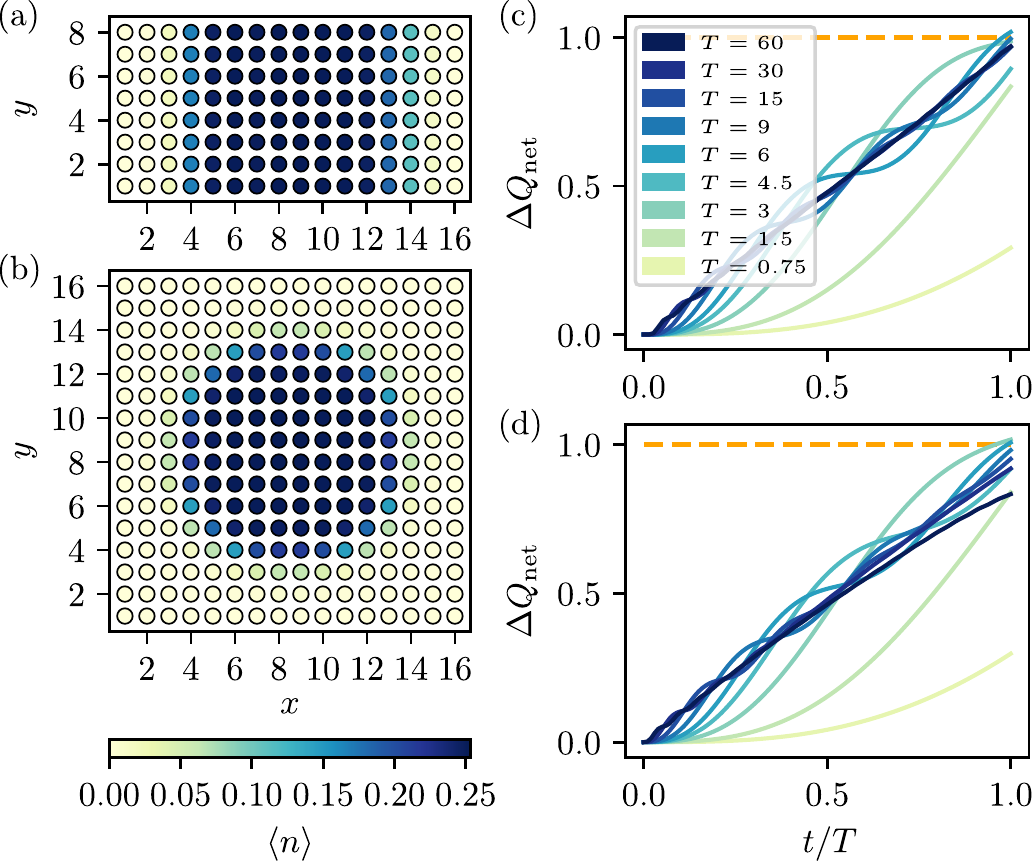}
    \caption{Comparison of the integer Chern insulator (CI) of free fermions on the cylinder and open geometry. 
    The particle densities depicted for the cylinder in (a) and the open system in (b) both show a wide region of $\expval{n} \approx 1/4$ corresponding to a completely filled lowest band with Chern number $C=1$. Figures (c) and (d) show the data for the pumped charge $\Delta Q_{\rm net}(T)$ for cylinder and open system, respectively. Their behaviors are qualitatively similar to the ones of the FCI system in Figs.~\ref{fig:pump}(d) and \ref{fig:pump_open}.}
    \label{fig:ff}
\end{figure}
We choose a cylinder system of $L_y=8$ with $N=20$ particles and $k_x=0.024$ and an open geometry of  $L_x \times L_y = 16 \times 16$ with $N=22$ particles and $k_x=k_y=0.024$. The density distribution of the ground states is shown in Fig.~\ref{fig:ff}(a) and \ref{fig:ff}(b) and the bulk of both systems displays a filling of $\expval{n} \approx 0.25$ corresponding to a CI in the lowest band. In the cylinder system, we create an electric field by threading $2\pi$ flux through the system during time $T$ resulting in an electric field of $E_y = -\pi/(4T)$ and monitor the value of $\Delta Q_{\rm net}$ defined in Eq.~\eqref{eq:pump} which is depicted in Fig.~\ref{fig:ff}(c). For the open case, we switch on a potential that causes a field of the same strength and we again average $\Delta Q_{i \to i-1}$ over the central four sites as in the open FCI case. Additionally, we correct by a factor of $4/5$ since the bulk part of the cloud extends over ten sites along the $y$ direction as opposed to eight on the cylinder. The pumped charge $\Delta Q_{\rm net}$ can be written dependent on $L_x$ as
\begin{equation}
\Delta Q_{\rm net} (t) = \frac {256}{5L_x^2} \sum_{i=x^-}^{x^+} \Delta Q_{i \to i-1} (t), \label{eq:pump_ff}
\end{equation} 
where $x^\pm = L_x/2 + \frac{1}{2} \pm (L_x/8 - \frac{1}{2})$.
As in the interacting system, the pumped charge in the open case shows a very similar behavior to the cylinder data for small to moderate $T$, but the response decreases again if $T$ becomes larger. 
As mentioned previously, the origin of this decrease is most likely due to the fact that the ratio between the electric field and the confining potential along the $y$ direction is becoming too small. 

\begin{figure}[t]
    \centering
    \includegraphics[width=\columnwidth]{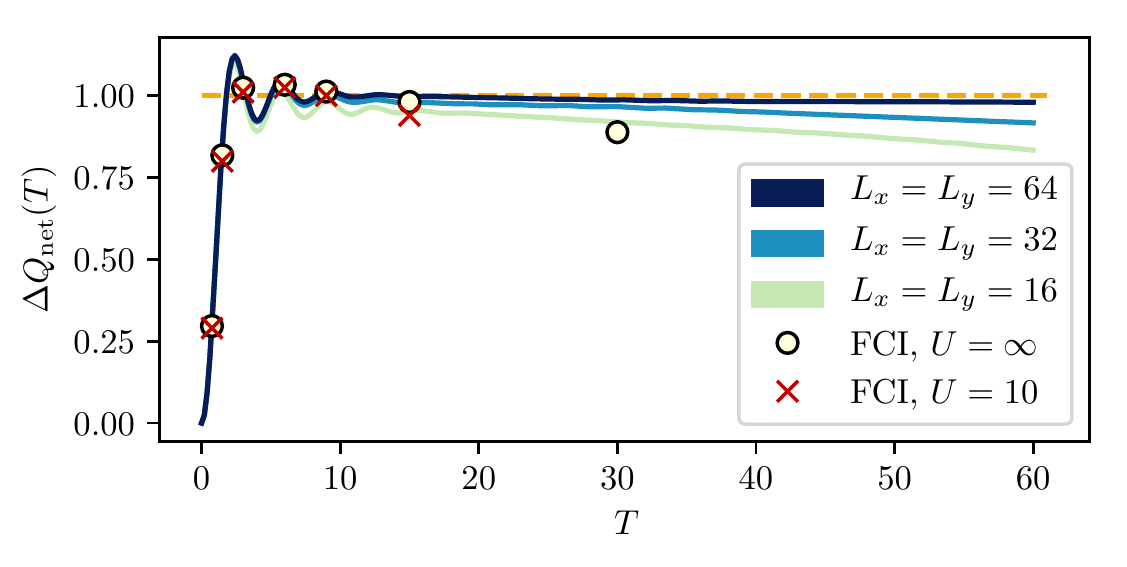}
    \caption{The charge pumping  as a function of total evolution time $T$ for the free fermion CI at three different system sizes (solid lines) and the FCI data from Fig.~\ref{fig:pump_open} (dots) for comparison. Small times exhibit clear nonadiabatic behavior with a value much smaller than unity. The value of $\Delta Q_{\rm net}$ decreases for large times, but this downturn is less pronounced for larger system size. The FCI data qualitatively agrees with that of its noninteracting counterpart. Additionally, we show FCI data for a finite interaction of $U=10$ which limits the  heating rate in experiments.}
    \label{fig:ff_scale}
\end{figure}

In order to gain further insight into the behavior at large times, we study the size dependence of the displacement since in larger systems, the confining potential strengths can be reduced. We consider two more setups with $L_x=L_y = 32, 64$, $N=88,352$ and $k_x=k_y= 0.006, 0.0015$. This sequence means we double the linear system size, multiply the particle number by four and divide the confinement strength by four for the subsequent size which ensures that the trapping potential has the same value around the edge of the particle cloud. 
Note that we do not alter the field strength between different sizes, the expression of $\Delta Q_{\rm net}$ in Eq.~\eqref{eq:pump_ff} is normalized such that it should approach unity at time $T$ if $\sigma_{xy}=1$. 
The data displayed in Fig.~\ref{fig:ff_scale} shows a clear nonadiabatic behavior for $T \lesssim 2 $ followed by oscillations around the quantized value and a downturn towards larger times. This downturn, however, softens as the system size grows so that a decent value of quantization ($0.98$ at $T=60$ for $L=64$) can be read off even for long times in the largest system. Additionally, we show the data of the FCI from Fig.~\ref{fig:pump_open} which qualitatively agrees with the free fermion system. Therefore, we expect that the downturn will be less prominent as well in larger interacting systems that could be realized experimentally, but are too expensive to simulate numerically. Even in the small FCI system that we evaluated, the data shows a reasonable quantization ($1 \pm 0.033$) in the time window of $T \sim 5 - 15$. 

Major concerns about the realization and detection of FCIs in cold atomic systems are effects of finite temperature and heating. Even if low enough temperatures to realize FCIs can be reached, the periodic driving to engineer the topological band structure leads to additional heating effects. Recently, it has been shown that there may be an ideal driving frequency window in which intraband heating is delayed due to prethermalization while at the same time an effective one-band description of the optical lattice is still justified \cite{Sun2020}. The coupling to higher bands also limits the interaction strength, but the stability of a $\nu=1/2$ Laughlin state has been demonstrated in fully time-dependent calculations \cite{Hudomal2019}. Therefore, in addition, we perform the simulation for the FCI in the open system for a finite realistic interaction of $U=10$. The results are depicted in Fig.~\ref{fig:ff_scale} demonstrating an agreement with the hardcore case for small $T$ and showing that the quantization can still be read off for $T \sim 5 - 10$ in this small system.

\textit{Conclusion}.---The Hall conductivity of a $\nu=1/2$ FCI confined by a harmonic potential can be accurately determined by measuring the transverse displacement of the particle cloud under the application of a constant force, both in a cylinder and open geometry. The force field has to be weak enough to avoid nonlinear effects, yet too small fields yield to a decrease of the signal in the open system. While cylindrical optical lattices including a dynamical flux threading can, in principle, be engineered \cite{Lacki2016}, the effect of the decrease in the displacement for small fields in the open case is weakening with increasing system size as indicated by the comparison to a noninteracting CI.

The harmonic confinement accounts for a realistic experimental setup and the displacement of the particles can be measured by taking snapshots of the particle cloud~\cite{Bakr2009}. 
The difference in position in our simulations is of the order of one lattice spacing which should be large enough to resolve experimentally. However, longer evolution times will linearly increase the displacement leading to a clearer experimental signal. As a comparison, the drift to determine the Chern number in Ref.~\cite{Aidelsburger2015} was measured over a time of $4/J$, comparable to the times at which the quantization is observable in our simulations. A clear signature at such small times is especially encouraging for experiments, as the FCI state will have a finite lifetime due to heating processes.

While completing this mansucript we became aware of a related work studying the Hall drift of an FCI in the interacting Harper-Hofstadter model \cite{Repellin2020}.

\textit{Acknowledgements}.---We thank C. Repellin, J. L\'eonard and N. Goldman for sharing their manuscript~\cite{Repellin2020} before submission and are grateful to A. Grushin, C. Repellin and S. Chatterjee for critical reading of an earlier version of our manuscript.
J.M. received funding through DFG research fellowship MO 3278/1-1 and TIMES at Lawrence Berkeley National Laboratory supported by the U.S. Department of Energy, Office of Basic Energy Sciences, Division of Materials Sciences and Engineering, under Contract No.\ DE-AC02-76SF00515. I.N. acknowledges support from the SK Hynix research fellowship and Kwanjeong Educational Foundation. Computations were performed on the Savio cluster at UC Berkeley and the Lawrencium platform at Lawrence Berkeley National Laboratory using the TenPy library~\cite{Kjall2013}.

\bigskip

\onecolumngrid
\newpage



\section*{Supplementary material}

# Supplemental Material for: Detecting Fractional Chern Insulators in Optical Lattices through Quantized Displacement

Johannes Motruk[1,2] and Ilyoun Na[1]

[1]*Department of Physics, University of California, Berkeley, California 94720, USA*
[2]*Materials Sciences Division, Lawrence Berkeley National Laboratory, Berkeley, California, 94720, USA*

(Dated: November 21, 2020)

## GROUND STATE PROPERTIES

Here, we give further details on the ground states in our systems, especially regarding their dependence on the strength of the confinement potential in the cylinder case. To check the parameter region of $k_x$ in which the FCI on the cylinder is stable, we compute the particle number in the central $8 \times L_y$ sites of the system as

$$N_c = \sum_{x=5}^{12} \sum_{y=1}^{L_y} \langle n_{x,y} \rangle. \quad (1)$$

This data is shown in Figs. S1(a)-(c). For all circumferences, we observe a wide plateau of $N_c \approx L_y$ for a range of $k_x$ values, signaling that the average filling is close to $1/8$. We expect that all $k_x$ values that show the corresponding particle number stabilize the FCI ($0.015 \lesssim k_x \lesssim 0.05$). To obtain a more detailed understanding of the density distribution, we focus on one parameter point deep inside the plateau at $k_x = 0.024$ and show the average density per site $N(x)$ as a function of $x$ in Figs. S1(e)-(f). It is defined as

$$N(x) = \frac{1}{L_y} \sum_{y=1}^{L_y} \langle n_{x,y} \rangle = \langle n_{x,y} \rangle. \quad (2)$$

The last equality follows from the translational invariance in the $y$-direction which implies that the expectation value at one site is equal to the average around the cylinder at fixed $x$. The density profile for $L_y = 4$ in Fig. S1(a) shows a clear charge density wave (CDW) pattern in the central region. This is consistent with the CDW order of the Tao-Thouless state arising in the thin torus limit of the FQH effect [1–4] which is adiabatically connected to the Laughlin state in two dimensions. Due to the interplay of filling factor, trapping potential and CDW order, the state at $L_y = 4$ breaks the reflection symmetry about the central bond of the system. We verified the existence of a nearly degenerate state with the CDW pattern shifted by one site again consistent with the thin torus limit state. The magnitude of the CDW order parameter is expected to decrease exponentially with the cylinder circumference [5]. Indeed, the data for $L_y = 6$ in Fig. S1(b) shows a reduced density modulation albeit the state still breaking the reflection symmetry. Additionally, the CDW pattern is less regular as the finite size of the

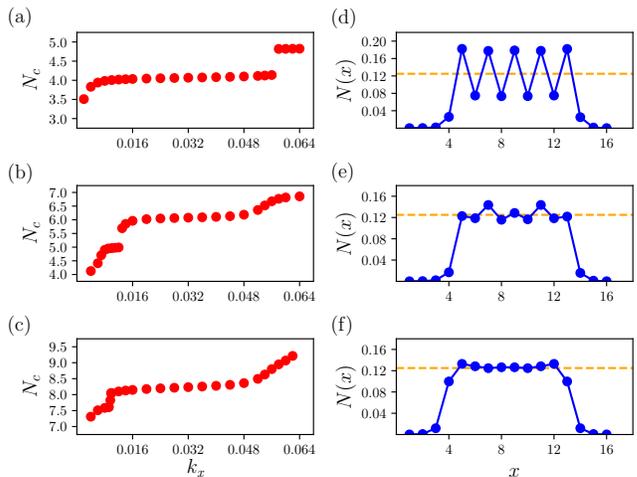

FIG. S1. (a)-(c) Number of particles $N_c$ in the central $8 \times L_y$ sites as a function of the confining strength $k_x$. For all three circumferences, a wide plateau of $N_c \approx L_y$ emerges which points to an average density of $\langle n \rangle \approx 1/8$ corresponding to a $\nu = 1/2$ filling of the lowest band. (d)-(f) Average particle density $N(x)$ at site $(x,y)$ for confining strength $k_x = 0.024$. This is the $k_x$ we focus on to compute the displacement in the main text. The $L_y = 4$ system shows a clear CDW pattern which decreases with increasing $L_y$, yet the average filling in the center of the system is close to $\langle n \rangle = 1/8$ for all system sizes indicated by the orange dashed line.

particle cloud influences it more strongly. At $L_y = 8$ in Fig. S1(c), there is no CDW-like density modulation observable and the reflection symmetry is restored indicating the crossover to 2D behavior. Due to boundary effects, it is not possible to define a meaningful CDW order parameter to check its exponential decay as a function of circumference, but the data qualitatively agrees with the theoretical picture. In Fig. S2, we show the density along a cut through the center of the system with square geometry considered in the main text with $L_x = L_y = 14$ and $k_x = k_y = 0.024$ where we also observe an extended central region of filling $1/8$.

The stationary particle currents in the ground state depicted in Fig. 2 of the main text are computed as follows. The current in the $x$-direction flowing from site $(x,y)$ to site $(x+1,y)$ is given by

$$\mathcal{J}^x_{(x \to x+1, y)} = \left\langle -iJ b^\dagger_{x,y} b_{x+1,y} + \text{H.c} \right\rangle \quad (3)$$

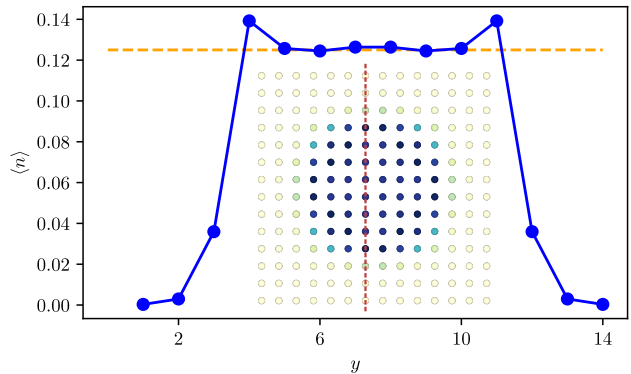

FIG. S2. Particle density along a cut through the center of the open system (see inset) displaying a central region of filling $\langle n \rangle \approx 1/8$.

## DETAILS OF MPS SIMULATIONS

We compute the ground state for the cylinder sizes and the open system with standard DMRG and ensure that the particle density has converged in the MPS bond dimension $\chi$. The displacement is then calculated using the matrix-product operator (MPO) time evolution algorithm introduced in Ref. [6] with the $W^{II}$ expression for the time evolution MPO matrices. In order to guarantee the reliability of our results, we perform the time evolution for MPS bond dimensions of $\chi = 100, 300$ and $500$ and time steps of $dt = 0.005, 0.01$ and $0.05$. We find that the results for the particle density and hence the value of the displacement do not change anymore in the cylinder systems between computations at $\chi = 300$ and $\chi = 500$ as well as between the choice of time steps $dt = 0.01$ and $dt = 0.005$ proving that these quantities have converged. The square geometry required to increase the bond dimension to $\chi = 800$ to guarantee convergence.

and the current in the $y$-direction flowing from site $(x, y)$ to site $(x, y+1)$ is defined as

$$\mathcal{J}^y_{(x,y \to y+1)} = \left\langle -iJe^{-i(\frac{\pi x}{2} - \frac{1}{L_y}\phi_\text{ext})} b^\dagger_{x,y} b_{x,y+1} + \text{H.c} \right\rangle \quad (4)$$

where $y = L_y + 1$ corresponds to $y = 1$. We clearly observe counterpropagating currents in $y$-direction around the circumference of the cylinder at the edges of the particle cloud showing the chiral nature of the state as a further indicator for the presence of the FCI. A representative current distribution for the cylinder case is shown in the upper density plot of Fig. 1 in the main text for the $L_y = 8$ system. For the square geometry, the chiral currents flow in clockwise direction around the particle cloud which is depicted in the lower plot of Fig. 2.



\end{document}